\documentclass[twocolumn,showpacs,amsmath,prd,amssymb,superscriptaddress,preprintnumbers]{revtex4}

\pdfoutput=1

\usepackage{graphicx}
\usepackage{dcolumn}
\usepackage{bm}

\def\nuA{\nu {\rm A}_{el}}
\def\q2{q^2}
\def\Enu{{E_{\nu}}}
\def\T0{T_{min}}
\def\sigmanuA{\sigma_{\nuA}}
\def\nue{\nu_e}
\def\nuebar{\bar{\nu}_e}
\def\numu{\nu_{\mu}}
\def\numubar{\bar{\nu}_{\mu}}

\begin{document}

\preprint{NCTS-ECP/1503}

\title{
Coherency in Neutrino-Nucleus 
Elastic Scattering
}

%

\newcommand{\as}{Institute of Physics, Academia Sinica,
Taipei 11529, Taiwan.}
\newcommand{\deu}{Department of Physics,
Dokuz Eyl\"{u}l University, Buca, \.{I}zmir 35160, Turkey.}
\newcommand{\bhu}{Department of Physics, Banaras Hindu University,
Varanasi 221005, India.}
\newcommand{\thu}{Department of Engineering Physics, 
Tsinghua University, Beijing 100084, China.}
\newcommand{\scu}{
College of Physical Science and Technology, Sichuan University, 
Chengdu 610064, China.} 
\newcommand{\ntu}{
Department of Physics, CTS and LeCosPA, National Taiwan University,
Taipei 10617, Taiwan.}
\newcommand{\ndhu}{
Department of Physics, National Dong Hwa University,
Shoufeng, Hualien 97401, Taiwan.}
\newcommand{\corr}{htwong@phys.sinica.edu.tw}

\author{ S.~Kerman } \affiliation{ \as } \affiliation{ \deu }
\author{ V.~Sharma }  \affiliation{ \as } \affiliation{ \bhu }
\author{ M.~Deniz } \affiliation{ \as } \affiliation{ \deu }
\author{ H.T.~Wong } \altaffiliation[Corresponding Author: ]{ \corr } \affiliation{ \as }
\author{ J.-W.~Chen }  \affiliation{ \ntu } 
\author{ H.B.~Li }  \affiliation{ \as } 
\author{ S.T.~Lin }  \affiliation{ \scu } 
\author{ C.-P.~Liu }  \affiliation{ \ndhu } 
\author{ Q.~Yue }  \affiliation{ \thu } 

\collaboration{TEXONO Collaboration}



\date{\today}

\begin{abstract}

Neutrino-nucleus elastic scattering provides 
a unique laboratory
to study the quantum mechanical coherency effects in 
electroweak interactions, towards which 
several experimental programs are 
being actively pursued.
We report results of our quantitative studies on
the transitions towards decoherency.
A parameter ($\alpha$) is identified to describe 
the degree of coherency, 
and its variations with incoming neutrino energy,
detector threshold and target nucleus are studied.
The ranges of $\alpha$ which can be probed 
with realistic neutrino experiments are derived,
indicating complementarity between projects with
different sources and targets.
Uncertainties in nuclear physics and in $\alpha$ 
would constrain sensitivities in 
probing physics beyond the standard model.
The maximum neutrino energies 
corresponding to $\alpha$$>$0.95 
are derived.

\end{abstract}


\maketitle


The elastic scattering of a neutrino 
with a nucleus~\cite{nuA-early,nuA-komas}
\begin{equation}
\nuA :
~~~~~~
\nu ~ + ~ A(Z,N) ~ \rightarrow ~
\nu ~ + ~ A(Z,N) ~~ ,
\label{eq::nuA}
\end{equation}
where $A(Z,N)$ denotes the atomic nucleus 
with its respective atomic, charge and
neutron numbers,
is a fundamental electroweak neutral current process
in the Standard Model (SM)
which has never been experimentally observed.
It can provide a sensitive probe to 
physics beyond SM (BSM)~\cite{nuA-BSM,scholberg}
and plays an important role in 
astrophysical processes~\cite{nuA-early,nuA-astro}.
It offers prospects to 
study neutron density distributions~\cite{nuA-ndd},
to detect supernova neutrinos~\cite{nuA-SN}
and to provide a compact and transportable
neutrino detector for real-time 
monitoring of nuclear reactors~\cite{nuA-monitor}.
The $\nuA$ events from
solar and atmospheric neutrinos are 
the irreducible background~\cite{nuA-DMfloor}
to forthcoming generation of 
dark matter experiments~\cite{rppdarkmatter}.
There are active experimental programs 
to observe and measure the processes 
with neutrinos from reactors~\cite{texononuA} or
from decay-at-rest pions (DAR-$\pi$)~\cite{scholberg} with 
a spallation neutron source~\cite{SNSnuA}.

The $\nuA$ reaction provides a unique laboratory
to study the quantum mechanical coherency effects 
in electroweak interactions.  At low momentum transfer,
the de Broglie wavelength of the neutrinos is large
compared with the nucleus, and the scattering 
amplitude of individual nucleons will coherently
add to contribute to the cross-sections.
Typically the neutrino energy ($\Enu$)
and the measurable nuclear recoil kinetic energy ($T$)
are much less than the target nucleus mass ($M$)
in the discussion of coherency.
The three-momentum transfer ($ q \equiv | \vec{q} |$)
is given by $\q2$=$2 M T + T^2$$\simeq$$ 2 M T$.
Kinematics places constraints to the maximum
recoil energy to be 
$T_{max}$=$2 \Enu ^2 / ( M + 2 \Enu )$$\simeq$$ 2 \Enu ^2 / M$.
 

\begin{figure}
{\bf (a)}\\
\includegraphics[width=8.2cm]{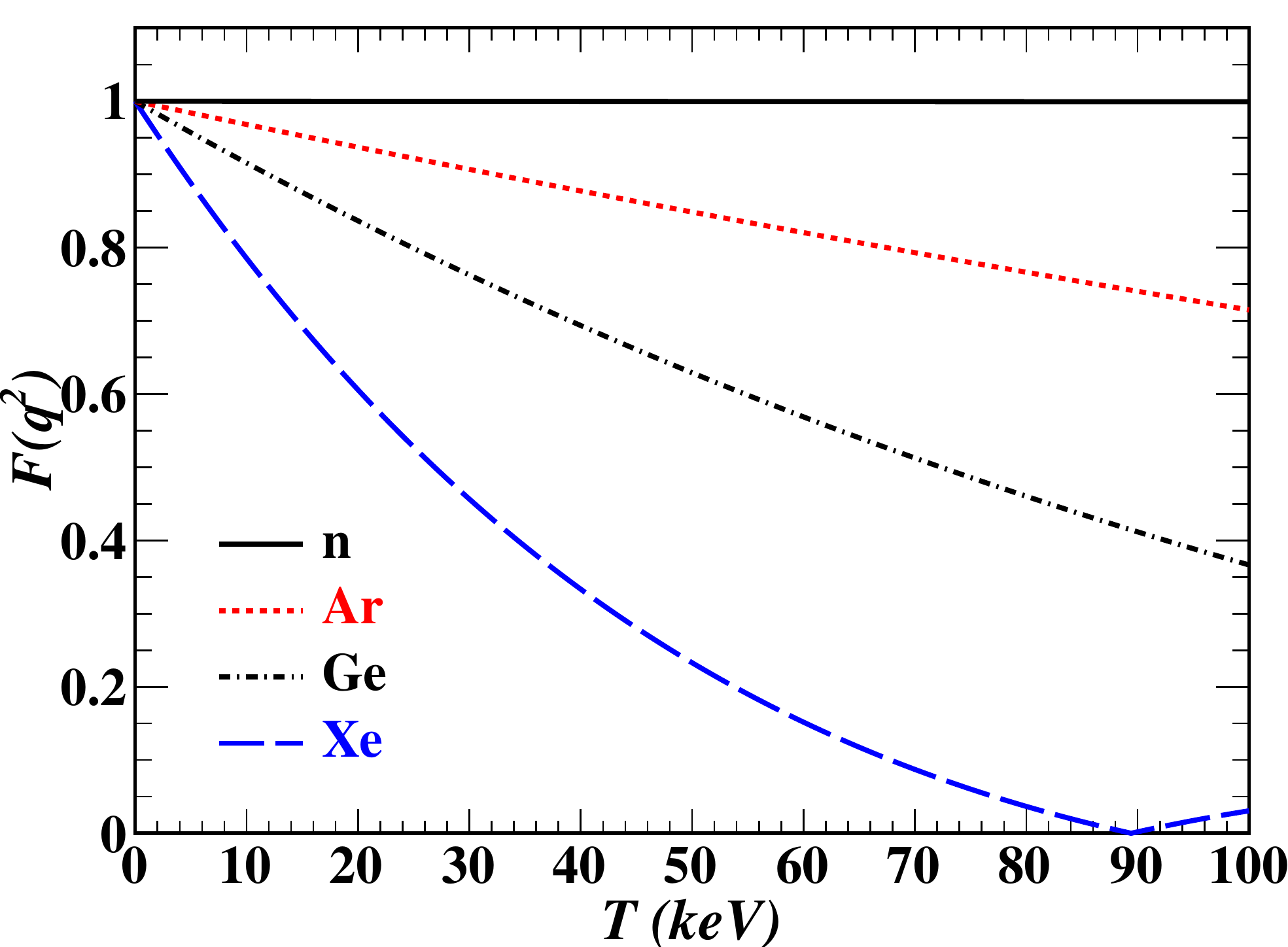}\\
{\bf (b)}\\
\includegraphics[width=8.2cm]{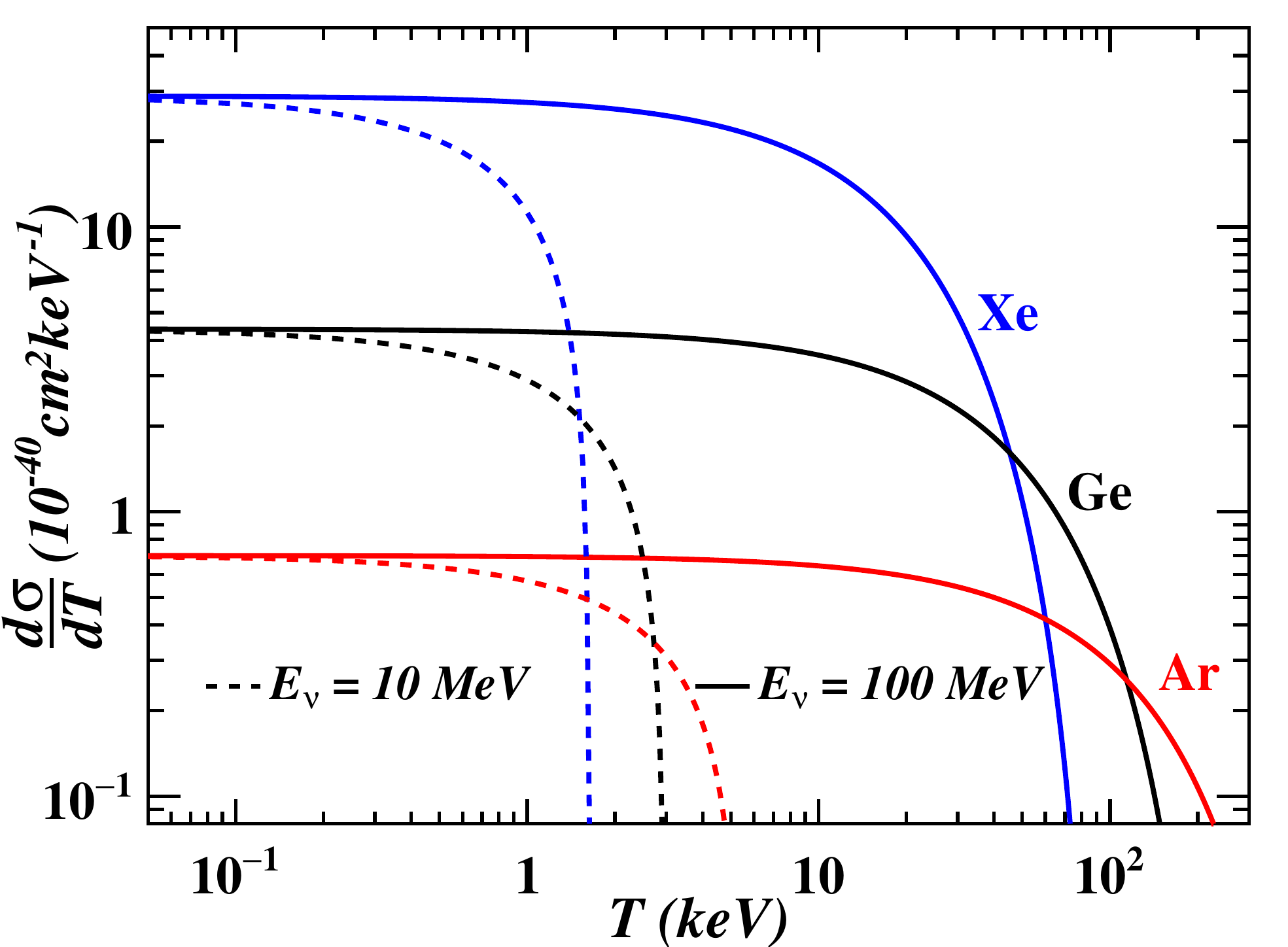}\\
{\bf (c)}\\
\includegraphics[width=8.2cm]{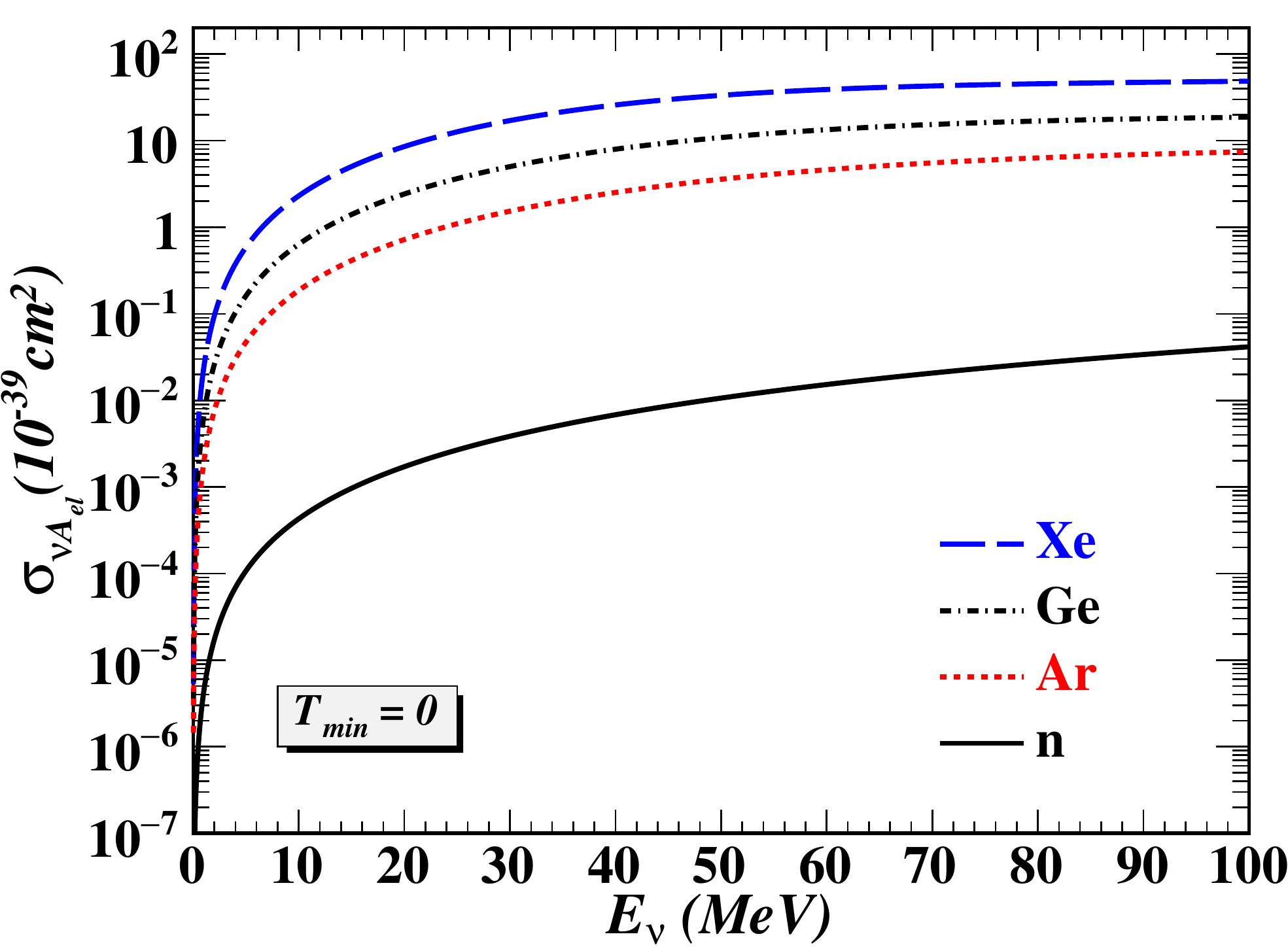}
\caption{
(a) 
nuclear form factor $F ( \q2 )$ as function of $T$,
related by $ \q2 = 2 M T$;
(b) 
differential cross-section of $\nuA$ as function
of $T$ at $\Enu$=(10,100)~MeV; and
(c) 
total cross-section ($\sigmanuA$) 
at $\T0$=0 as function of $\Enu$.
The nuclei (n,Ar,Ge,Xe) are selected 
for illustrations.
}
\label{fig::sigmanuA}
\end{figure}


A generic scale of $\Enu$$<$50~MeV 
is usually taken to characterize the 
requirement of coherency.
The theme of this article is to 
quantify this transition $-$
the first such investigation in the literature.
We parametrize the degree of coherency 
and study its dependence on $\Enu$,
nuclear recoil detection threshold ($\T0$),
and the nucleus $A(Z,N)$.
Potential reaches of the
up-coming experimental programs
are derived.


The differential cross-section of $\nuA$ scattering
in SM is given by~\cite{nuA-komas}: 
\begin{eqnarray}
\label{eq::diffsigmanuA}
\frac{d \sigmanuA}{d \q2} ( \q2 , \Enu )  &   =   &  
\frac{1}{2} ~ [ \frac{ G_F^2 }{ 4 \pi}  ] ~ 
[ 1 - \frac{ \q2 }{ 4 \Enu ^2 } ] \\ \nonumber
& & \left[ \varepsilon Z F_Z ( \q2 ) - N F_N ( \q2 ) \right] ^2 
\end{eqnarray}
or, equivalently, in the experimental measurable $T$ as 
\begin{equation}
\frac{d \sigmanuA}{d T} ~   =  ~    
 2 M  ~ \left[ \frac{d \sigmanuA}{d \q2} \right]
\label{eq::diffsigmanuAinT}
\end{equation}
where $F_Z ( \q2 )$ and $F_N ( \q2 )$ are, respectively,
the proton and neutron nuclear form factors
for $A(Z,N)$, while
$\varepsilon$$\equiv$(1$- 4 ~ {\rm sin^2 \theta_W } )$=0.045,
indicating the dominant contributions are
from the neutrons.
The total cross-section depends on
$( \Enu , \T0 ;  M , Z , N )$
and is given by:
\begin{equation}
\sigmanuA = 
\int_{\q2_{min}}^{\q2_{max}} 
\left[ \frac{d \sigmanuA }{d \q2} ( \q2 , \Enu )  \right] ~ d \q2
\label{eq::totalsigmanuA}
\end{equation}
where the integration limits of
$\q2_{max}$=$4 \Enu ^2 [ M / ( M$+$2 \Enu ) ]$$\simeq$$4 \Enu ^2$ and
$\q2_{min}$=$2 M \T0$ are defined by
the kinematics and detection threshold, respectively.
A threshold of $\T0$$<$$2 \Enu ^2 / M$ 
is required to detect
neutrinos of energy $\Enu$.


\begin{table}
\caption{
Illustrations at $\Enu$=50~MeV $-$
the maximum nuclear recoil energy, 
the corresponding lower bounds of $F ( \q2 )$, 
as well as the coherency factors $\alpha$ and $\xi$ 
at $\T0$=0.
}
\begin{ruledtabular}
\begin{tabular}{ccccc}
& n & Ar & Ge & Xe \\ \hline
$T_{max}$~(keV) & 4810 & 134 & 73.8 & 40.9 \\
$ F ( \q2)  >$ & 0.97 & 0.63 & 0.49 & 0.32 \\
$\alpha >$  & $-$ & 0.77 & 0.68 &  0.57 \\
$\xi >$  & $-$ & 0.78  & 0.69 & 0.58
\end{tabular}
\end{ruledtabular}
\label{tab::Enu50MeV}
\end{table}


Various aspects in the calculations
of the nuclear form factors 
of Eq.~\ref{eq::diffsigmanuA}
are recently
discussed in Ref.~\cite{nuA-komas}.
We adopt the effective method of Ref.~\cite{engel} 
which assumes the same form factors for neutrons
and protons:
$F_Z ( \q2 )$=$F_N ( \q2 )$$\equiv$$F ( \q2 )$$\in$[0,1], 
with
\begin{equation}
F ( \q2 ) =
[ \frac{3}{q R_0} ] ~
J_1 ( q R_0 )
~ {\rm exp}  [ - \frac{1}{2} \q2 s^2 ]  ~~~ ,
\label{eq::formfactor}
\end{equation}
where $J_1 ( x )$ is 
the first-order spherical Bessel function.
The target nuclei dependence is introduced through 
$R_0^2$=$R^2$$-$$5 s^2$,
$s$=0.5~fm and $R$=$1.2 A^{\frac{1}{3}} ~ {\rm fm}$.
An alternative derivation~\cite{nuA-ndd} gives
form factors consistent to $<$0.07(1.7)\%
within the kinematic ranges 
corresponding to $\Enu$=10(50)~MeV.

Several nuclei with experimental interest and having
different mass ranges 
$-$ (neutron,Ar,Ge,Xe) at $Z$=(0,18,32,54) $-$
are selected for studies. 
(CsI, having $Z$=55 and 53,
can be approximated as Xe in this discussion).
Their corresponding differential and 
total cross-sections are the averages of 
Eqs.~\ref{eq::diffsigmanuAinT}\&\ref{eq::totalsigmanuA} 
due to individual isotopes
weighted by their respective isotopic-abundances.
The nuclear form factors $F ( \q2 )$ as function of $T$,
the differential cross-sections
at fixed $\Enu$=(10,100)~MeV and
the total cross-sections at $\T0$=0 are
depicted in Figures~\ref{fig::sigmanuA}a,b\&c, respectively.
The $F ( \q2 )$ ranges at $\Enu$=50~MeV
are illustrated in Table~\ref{tab::Enu50MeV}. 
The nuclear effects as characterized by the
deviations from unity are significant for heavy nuclei. 

At $\q2$$\rightarrow$0 and $F ( \q2 )$$\simeq$1,
full coherency is achieved when
the scattering amplitudes due to individual nucleons
are perfectly aligned and are summed with
no relative phase angle, such that 
the total cross-section is maximal. 
In particular at $\T0$=0,
\begin{equation}
\sigmanuA ( \T0 = 0 ) =
\frac{ G_F^2 \Enu ^2 }{ 4 \pi }
\left[ \varepsilon  Z  -  N \right] ^2 ~~ .
\label{eq::sigma-T0}
\end{equation}
The experimental signature of full coherency
is that $\sigmanuA$ varies as $[ \varepsilon  Z$$-$$N ] ^2$.
Another feature is that the differential cross-section
at small T,
\begin{equation}
\frac{d \sigmanuA}{d T} ( T \rightarrow 0 )  \simeq
[ \frac{ G_F^2 M }{ 4 \pi}  ]
\left[ \varepsilon Z - N \right] ^2
\label{eq::dsigma/dT}
\end{equation}
varies with the same factor and
is independent of $\Enu$, as depicted in
Figure~\ref{fig::sigmanuA}b.


We note that ``coherent pion production''~\cite{cpp-overview} 
with accelerator neutrinos at high energy
is a distinctly different process.
The coherency is due to the coupling
of a virtual meson with the nucleus producing a 
physical pion, and hence is a strong interaction
effect and varies as $A^2$. 
The coherency in $\nuA$ interactions,
on the other hand, 
is due to the coupling of a virtual Z-boson with 
the nucleus, and hence is an electroweak process.



\begin{figure}
{\bf (a)}\\
\includegraphics[width=8.2cm]{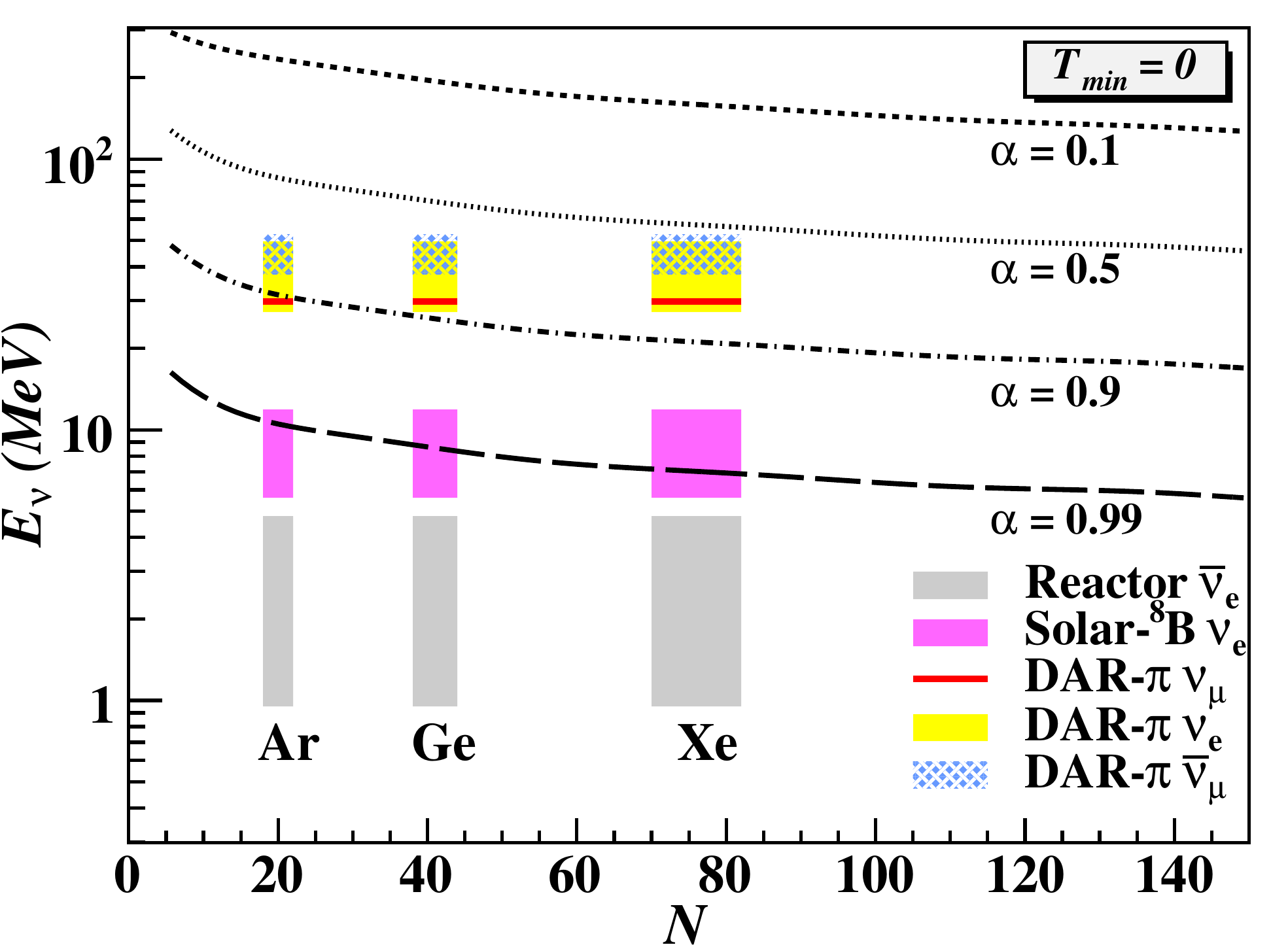}\\
{\bf (b)}\\
\includegraphics[width=8.2cm]{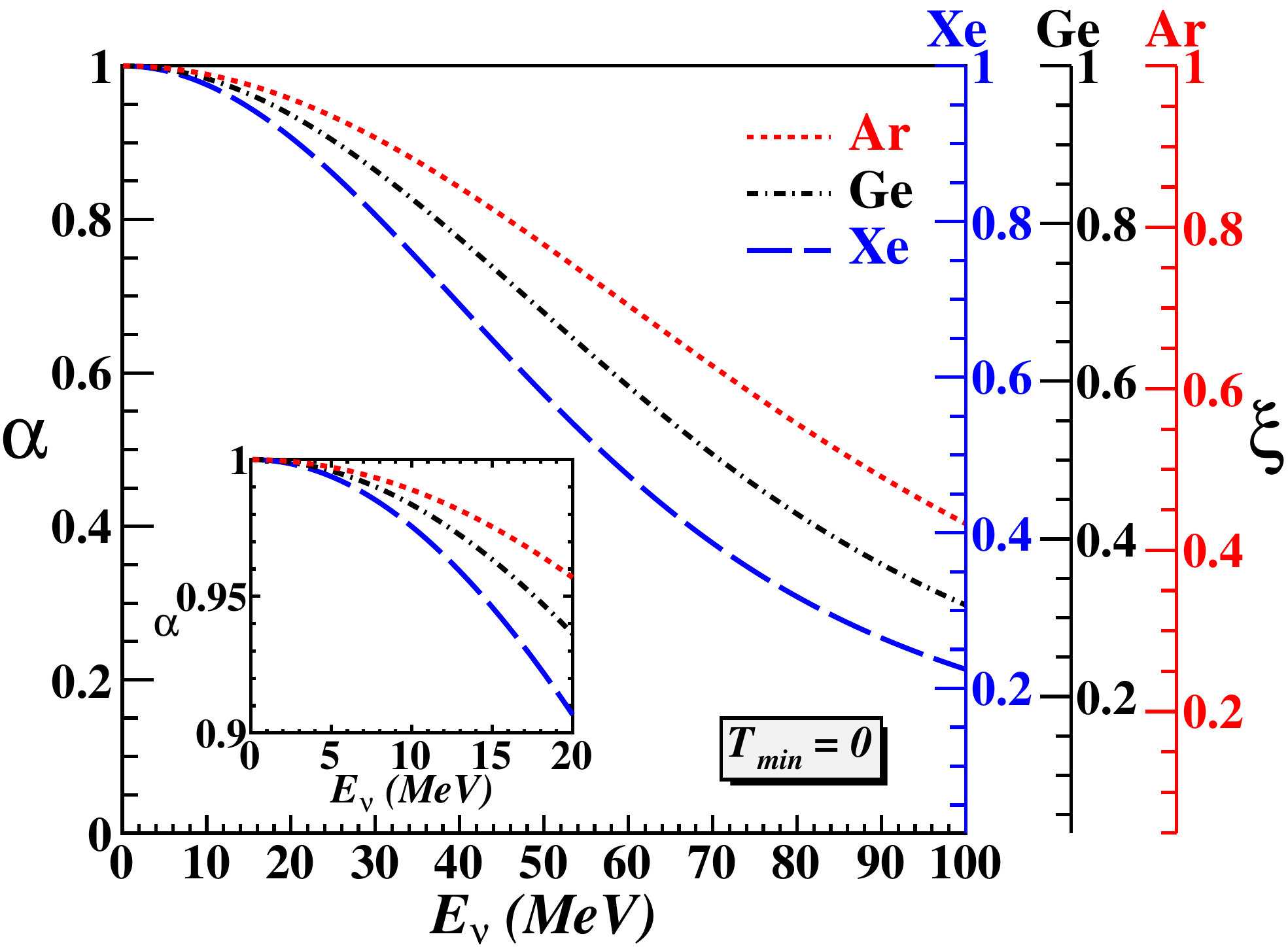}\\
{\bf (c)}\\
\includegraphics[width=8.2cm]{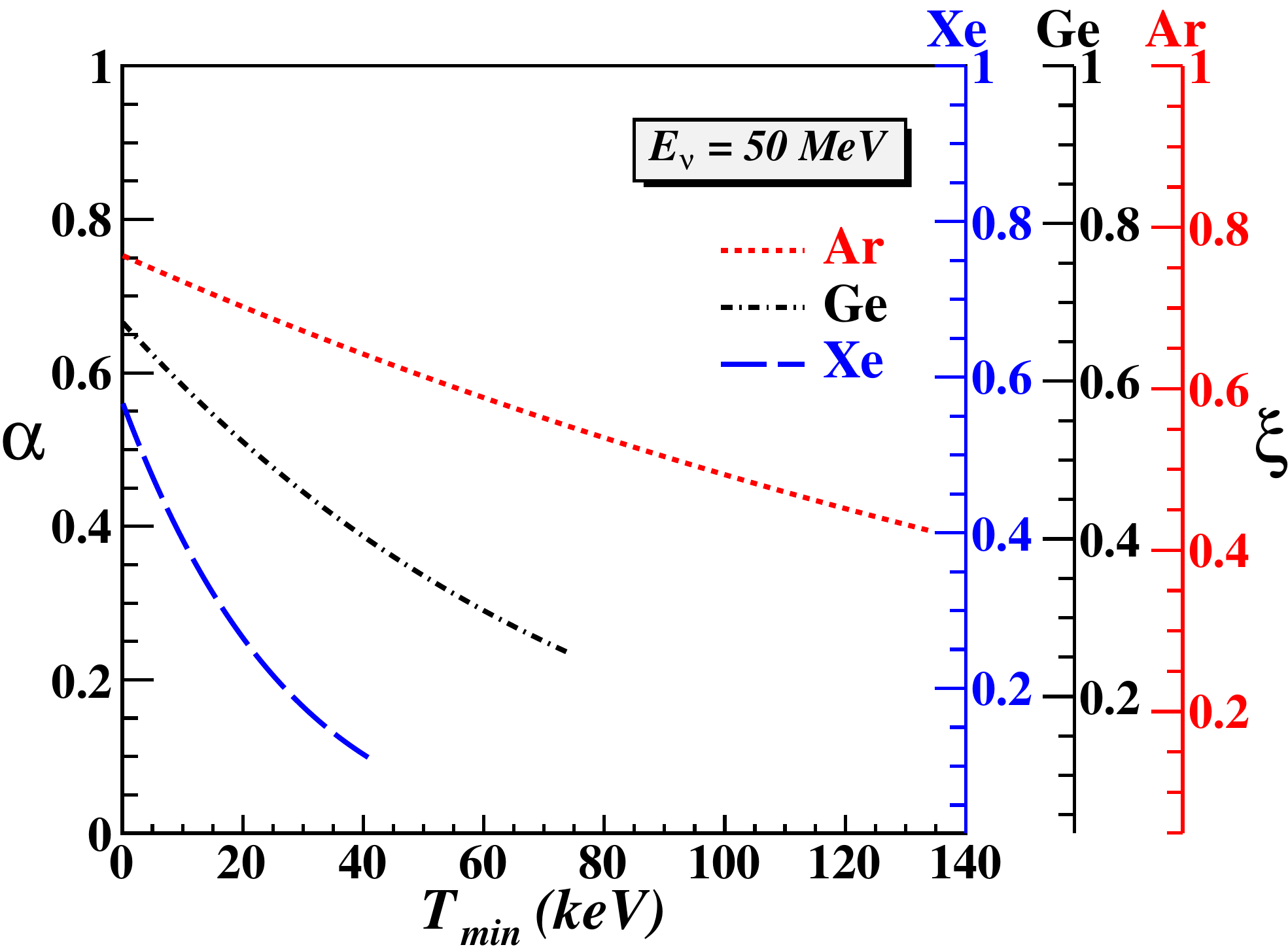}
\caption{
(a) The $\alpha$-contours on the
($N$,$\Enu$) plane at $\T0$=0,
with bands of realistic neutrino sources
and target nuclei superimposed.
Variations of $\alpha$ and $\xi$
for Ar,Ge,Xe as functions of
(b) $\Enu$ at $\T0$=0, and
(c) $\T0$  at $\Enu$=50~MeV,
where the end-points correspond to
maximum recoil energies.
}
\label{fig::alpha-Enu}
\end{figure}


\begin{figure}
{\bf (a)}\\
\includegraphics[width=7.8cm]{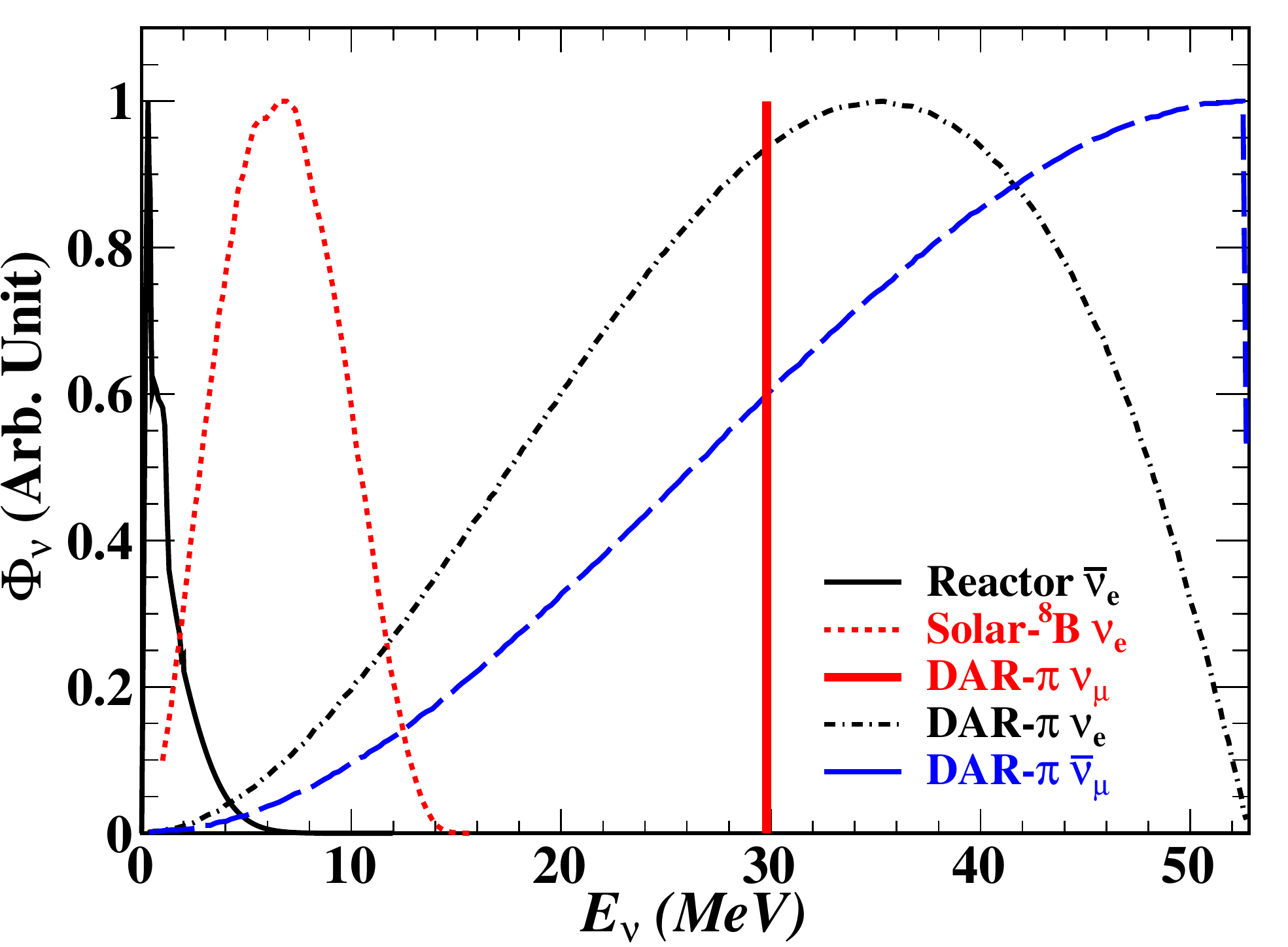}\\
{\bf (b)}\\
\includegraphics[width=7.8cm]{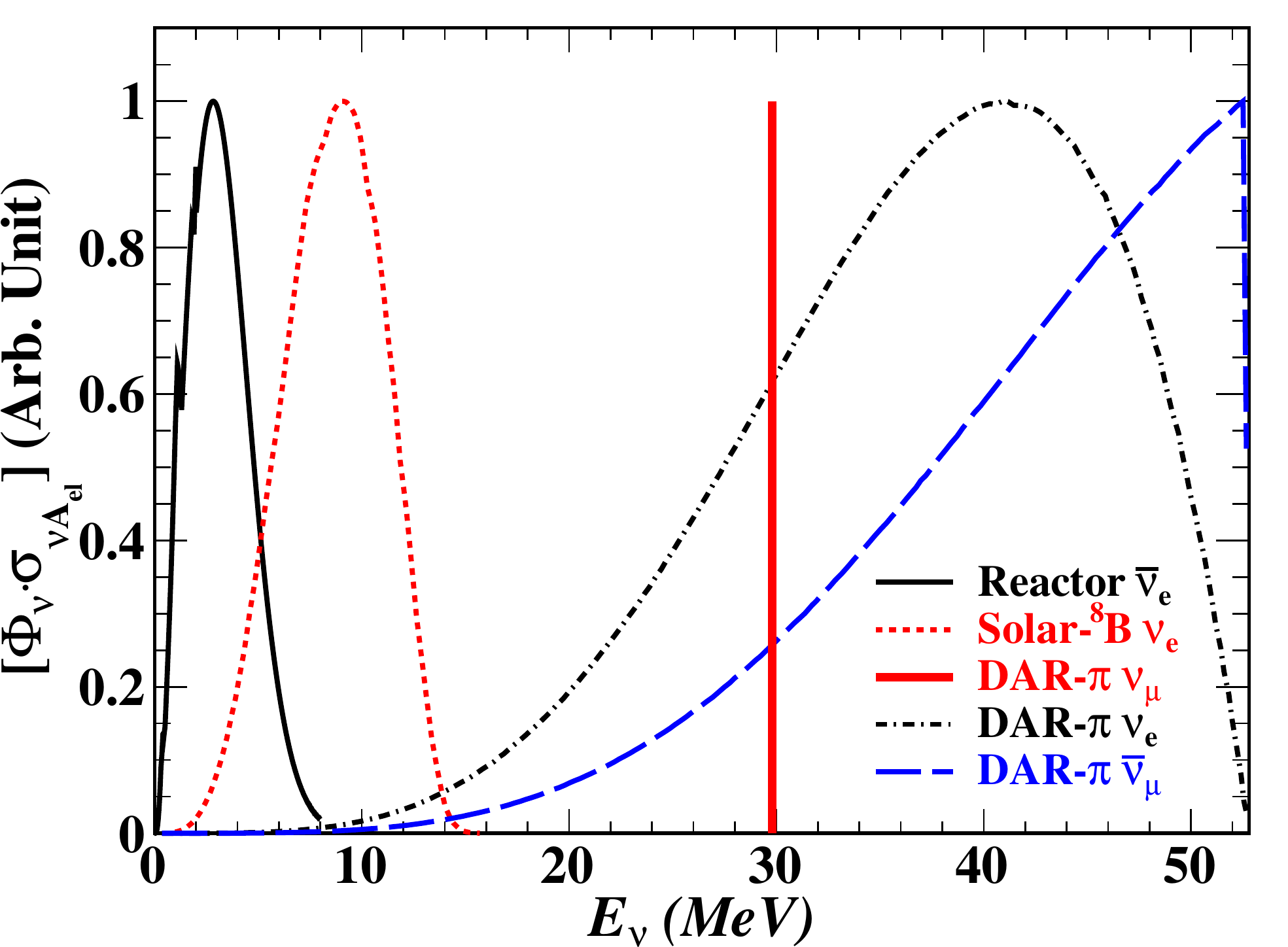}
\caption{
(a)
Neutrino spectra ($\Phi_{\nu}$)
from reactor $\nuebar$, 
DAR-$\pi$ ($\numu$,$\nue$,$\numubar$), and
solar-$^8$B $\nue$,
normalized by their maxima.
(b)
Distributions of 
[$\Phi_{\nu}$$\cdot$$\sigmanuA$] 
at $\T0$=0, which are the weights 
in the averaging of ($\alpha$,$\xi$) 
to provide measurements of
($\left< \alpha \right>$,$\left< \xi \right>$). 
}
\label{fig::nuspect}
\end{figure}


\begin{figure}
{\bf (a)}\\
\includegraphics[width=8.2cm]{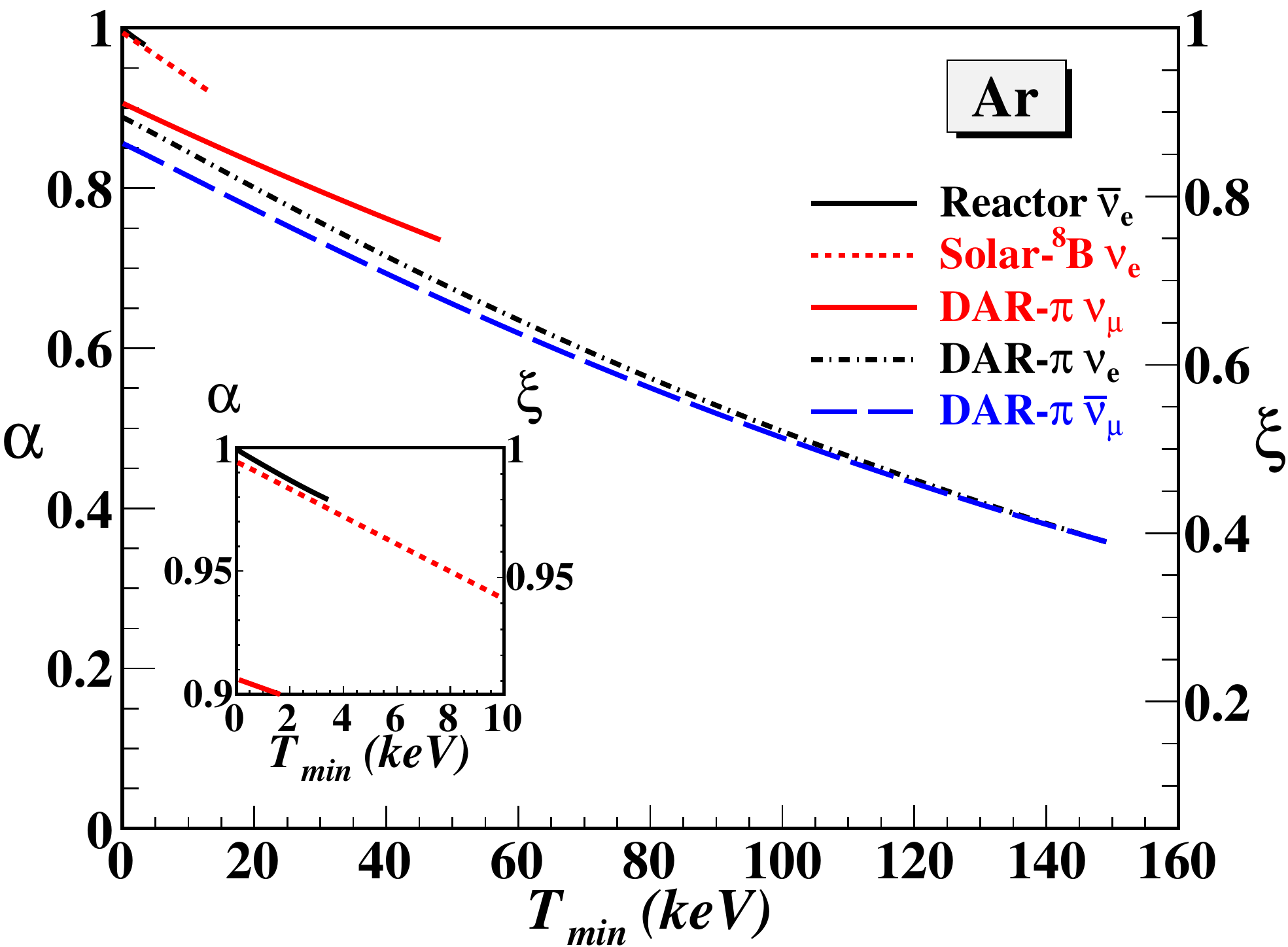}\\
{\bf (b)}\\
\includegraphics[width=8.2cm]{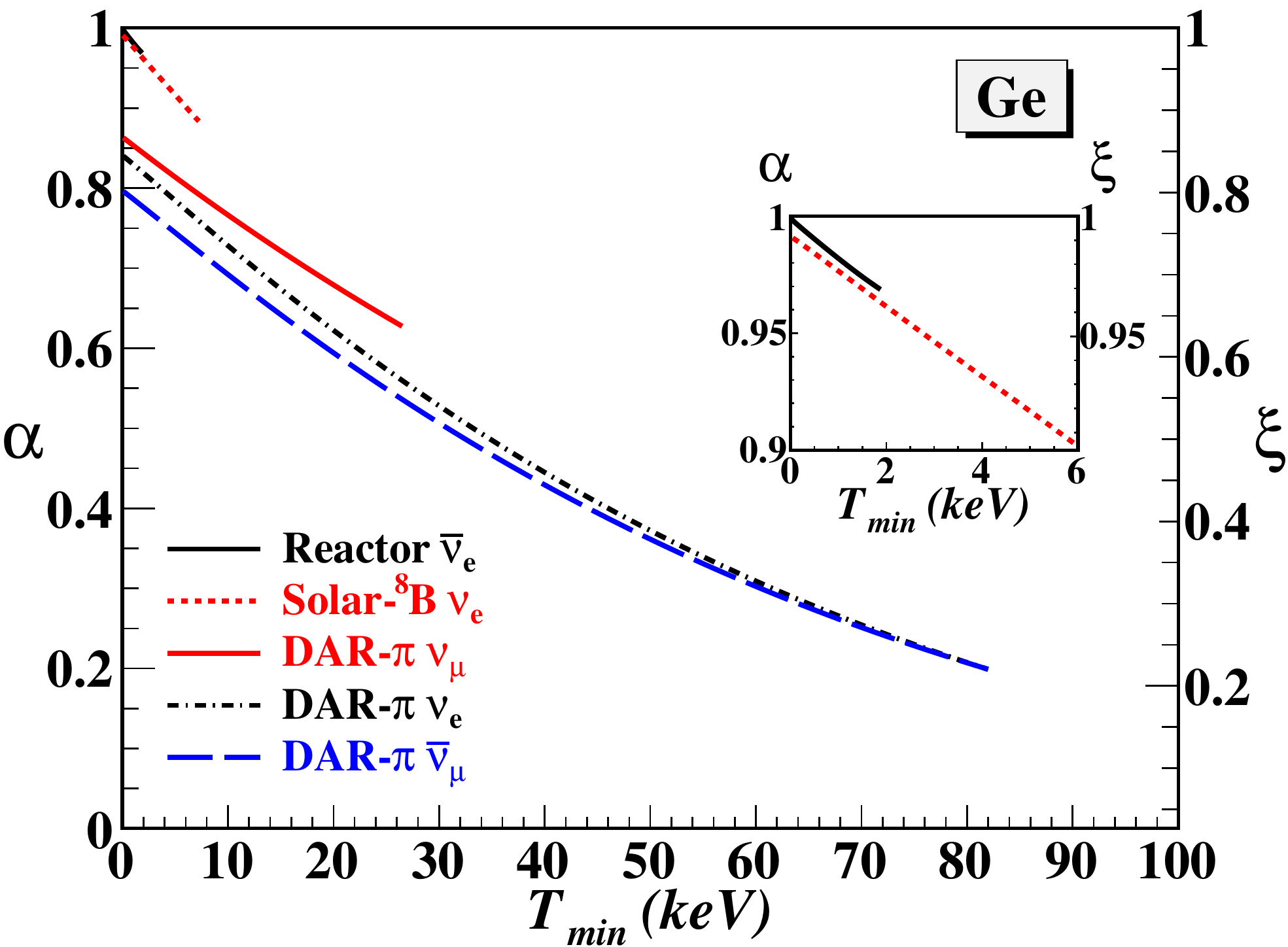}\\
{\bf (c)}\\
\includegraphics[width=8.2cm]{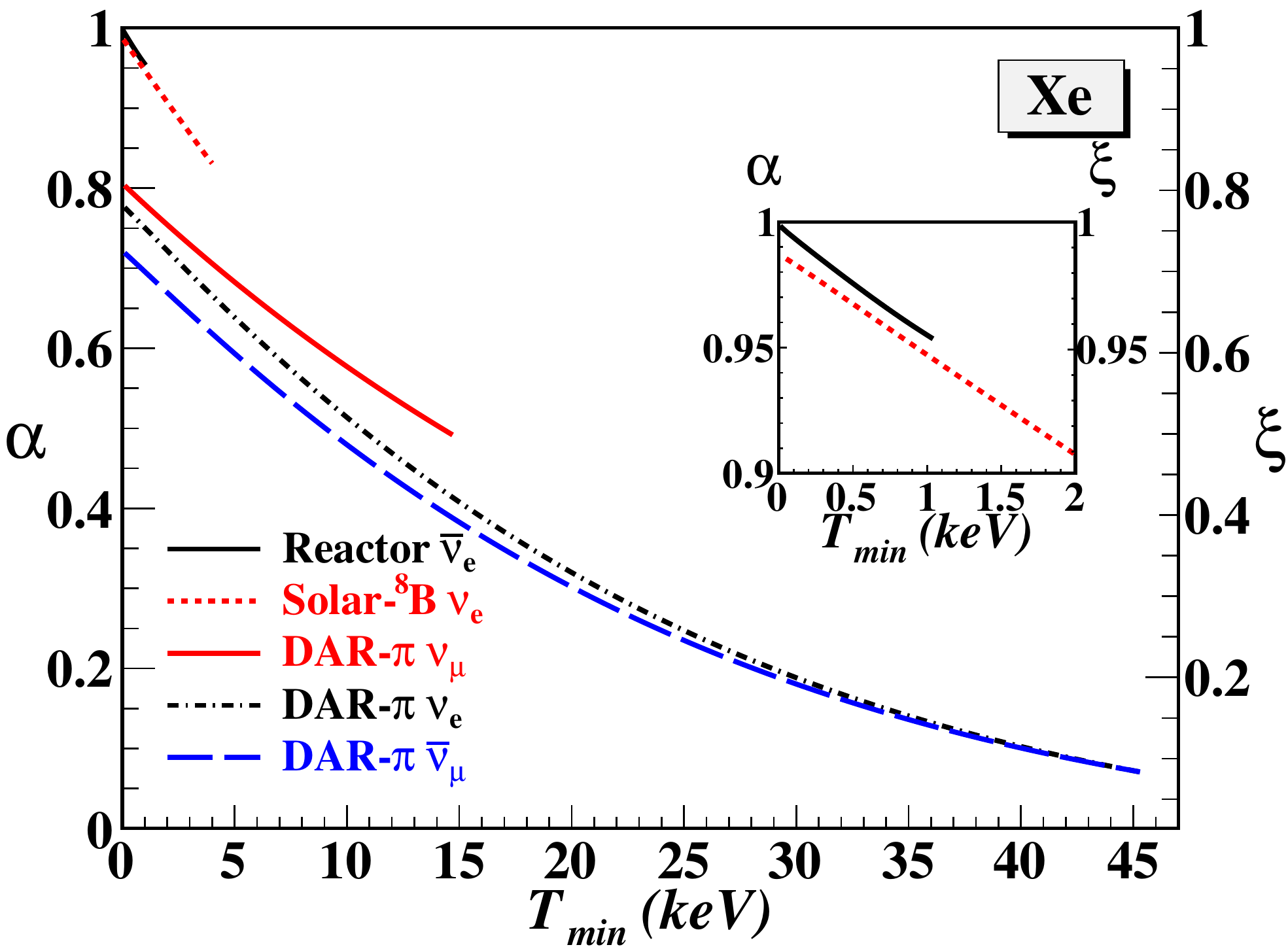}
\caption{
Variations of
($\left< \alpha \right>$,$\left< \xi \right>$)
as function of $\T0$
with the reactor $\nuebar$, solar-$^8$B $\nue$
and DAR-$\pi$ ($\numu$,$\nue$,$\numubar$),
for (a) Ar, (b) Ge, 
and (c) Xe.
The end-points correspond to maximum recoil
energies allowed by kinematics.
}
\label{fig::averaged-alpha}
\end{figure}


Departure from coherency for $\sigmanuA$
is characterized by 
deviations from the $[ \varepsilon  Z$$-$$N ] ^2$ scaling
as $\q2$ increases. 
The amplitude vectors of the different nucleons 
would add with a finite relative phase angle rather
than being perfectly aligned.
The combined amplitude $\mathcal{A}$
can be described by
\begin{equation}
\label{eq::amplitude}
\mathcal{A} ~  =  ~
\sum_{j=1}^{Z} e^{i \theta_j} \mathcal{X}_j + 
\sum_{k=1}^{N} e^{i \theta_k} \mathcal{Y}_k 
\end{equation}
where $\mathcal{X}_j ( \mathcal{Y}_k )$
denotes the coupling strength 
while $e^{i \theta_j} ( e^{i \theta_k})$
is the phase
for protons(neutrons).
For electroweak processes such as $\sigmanuA$, 
$( \mathcal{X}_j , \mathcal{Y}_k )$=$( - \varepsilon , 1 )$.

The cross-section comprises 
$(N+Z)^2$ terms: 
\begin{eqnarray}
\label{eq::formulation}
\sigmanuA ~ & \propto  & ~  \mathcal{A} ~ \mathcal{A}^\dagger ~~  \\
& = & ~
\sum_{j=1}^Z \mathcal{X}_j^2  ~ + ~
\sum_{k=1}^N \mathcal{Y}_k^2   \nonumber \\
& & + ~ \sum_{j=l+1}^{Z} \sum_{l = 1}^{Z-1} 
\left[ e^{ i ( \theta_j - \theta_l ) }  +
 e^{ - i ( \theta_j - \theta_l ) } \right]
\mathcal{X}_j \mathcal{X}_l  \nonumber \\
& & + ~ \sum_{k=m+1}^N \sum_{m = 1}^{N-1} 
\left[ e^{ i ( \theta_k - \theta_m ) }  +
 e^{ - i ( \theta_k - \theta_m ) } \right]
\mathcal{Y}_k \mathcal{Y}_m  \nonumber  \\
& & + ~ 
\sum_{j=1}^Z \sum_{k=1}^N
\left[ e^{ i ( \theta_j - \theta_k ) }  +
 e^{ - i ( \theta_j - \theta_k ) } \right]
\mathcal{X}_j \mathcal{Y}_k ~ . \nonumber  
\end{eqnarray}
Assuming that the decoherence effects 
between any nucleon pairs
can be described by
the average phase mis-alignment angle
$\left< \phi \right>$$\in$$[ 0 , \pi /2 ]$,
it follows that
\begin{equation}
\label{eq::alpha}
\left[ e^{ i ( \theta_j - \theta_k ) }  +
 e^{ - i ( \theta_j - \theta_k ) } \right]
=
2 ~ {\rm cos } ( \theta_j - \theta_k )  
=
2 ~ {\rm cos } \left< \phi \right> 
\end{equation}
and identically for
the other subscript pairs in Eq.~\ref{eq::formulation}  
with $(j,k)$$\leftrightarrow$$(j,l)$ and $(k,m)$.
The degree of coherency can therefore be 
quantified by a measurable parameter $\alpha$,
defined as 
$\alpha$$\equiv$${\rm cos }  \left< \phi \right>$$\in$$[0,1]$.

The cross-section ratio between 
$A(Z,N)$ and neutron(0,1),
following expansion of Eq.~\ref{eq::formulation}
and assignment of 
$( \mathcal{X}_j , \mathcal{Y}_k )$,
is given by:
\begin{eqnarray}
\label{eq::sigmaratio}
& & \frac{ \sigmanuA ( Z , N ) }{ \sigmanuA ( 0 , 1 )} \\
& = & 
 \{ ~ \varepsilon^2 Z + N 
+ \varepsilon^2 Z ( Z - 1 ) \alpha  
+ N ( N - 1 ) \alpha 
 - 2 \varepsilon Z N \alpha ~ \} \nonumber \\
& = & 
 \{ ~ Z \varepsilon ^ 2 [ 1 + \alpha ( Z - 1 ) ] +   
N [ 1 + \alpha ( N - 1 ) ]  
 -  2 \alpha \varepsilon Z N  ~ \} ~ . \nonumber 
\end{eqnarray}
The limiting conditions are: 
(a) $\alpha$=1 implies full coherency or
$\sigmanuA$$\propto$$[ \varepsilon  Z - N ] ^2$, 
while 
(b) $\alpha$=0 brings total decoherency or
$\sigmanuA$$\propto$$[ \varepsilon ^2  Z + N ]$.

The relative change in cross-section 
is an alternative parameter 
to characterize partial coherency:
\begin{equation}
\label{eq::xi}
\xi  \equiv  
\frac{ \sigmanuA ( \alpha ) }{ \sigmanuA ( \alpha = 1 ) } 
 =  \alpha +
( 1 - \alpha ) 
\left[   \frac{ ( \varepsilon ^2  Z + N ) }
{ ( \varepsilon  Z - N  ) ^2 } \right] ~~  .
\end{equation}
It readily follows that $\xi$ varies 
linearly with $\alpha$, and both are unity at
full coherency.

The parameters ($\alpha$,$\xi$) 
are evaluated with 
Eqs.~\ref{eq::sigmaratio}\&\ref{eq::xi}, respectively,
using form factors of Eq.~\ref{eq::formfactor}.
The $\alpha$-contours on the ($N$,$\Enu$) plane at $\T0$=0
are displayed in Figures~\ref{fig::alpha-Enu}a,
with the bands of realistic neutrino sources
and detector target nuclei superimposed.
The variations with $\Enu$ at $\T0$=0
are depicted in Figure~\ref{fig::alpha-Enu}b.
There is already significant decoherency 
at $\Enu$=50~MeV, with values listed 
in Table~\ref{tab::Enu50MeV}.
The coherency would further decrease with increasing
detector threshold, as illustrated in
Figure~\ref{fig::alpha-Enu}c. 


Experimental studies of coherency 
would be performed with realistic
neutrino sources. The current projects are based
on reactor $\nuebar$~\cite{texononuA}, 
DAR-$\pi$ ($\numu$,$\nue$,$\numubar$)~\cite{SNSnuA},
as well as the high energy solar-$^8$B $\nue$ in 
dark matter experiments~\cite{rppdarkmatter}.
These neutrino spectra ($\Phi_{\nu}$)~\cite{nuspectra,SNSnuA}
are depicted in Figure~\ref{fig::nuspect}a.
Experiments on $\nuA$ scattering provide measurements of 
($\left< \alpha \right>$,$\left< \xi \right>$), 
which are averages of ($\alpha$,$\xi$) weighted with
the distributions of [$\Phi_{\nu}$$\cdot$$\sigmanuA$]. 
Those at $\T0$=0 are displayed in Figure~\ref{fig::nuspect}b. 



\begin{table}
\caption{
The half-maxima in the
distributions of
[$\Phi_{\nu}$$\cdot$$\sigmanuA$] at $\T0$=0 
for the different neutrino sources,
and the values of $\left< \alpha \right>$ probed
by the selected target nuclei.
The $\numu$ from DAR-$\pi$ is mono-energetic.
}
\begin{ruledtabular}
\begin{tabular}{ccccc}
$\nu$ &  Half-Maxima of [$\Phi_{\nu}$$\cdot$$\sigmanuA$] &
\multicolumn{3}{c}{$\left< \alpha \right>$ ~ with}\\
Source  &  in $\Enu$(MeV)  &
Ar & Ge & Xe \\ \hline
Reactor $\nuebar$ & 0.96$-$4.82 & 1.00  & 1.00   & 1.00  \\
Solar-$^8$B $\nue$ & 5.6$-$11.9 & 0.99 & 0.99 & 0.98 \\
DAR-$\pi$ $\numu$ & 29.8 & 0.91 & 0.86 &  0.80 \\
DAR-$\pi$ $\nue$ & 27.3$-$49.8 & 0.89 & 0.83 & 0.76 \\
DAR-$\pi$ $\numubar$ & 37.5$-$52.6 & 0.85 & 0.79 & 0.71
\end{tabular}
\end{ruledtabular}
\label{tab::averaged-alpha}
\end{table}


The variations of
($\left< \alpha \right>$,$\left< \xi \right>$)
with detector threshold $\T0$ 
for the different neutrino sources 
and targets (Ar,Ge,Xe)
are depicted in
Figures~\ref{fig::averaged-alpha}a,b\&c, respectively. 
The values of $\left< \alpha \right>$ at $\T0$=0 are 
summarized in Table~\ref{tab::averaged-alpha}.
There is strong complementarity between
the configurations.
The combined measurements of differential cross-sections
allow studies of the transitions from coherency
to decoherency in $\nuA$.
In particular the low energy reactor $\nuebar$ and
solar-$^8$B $\nue$ probe the
full coherency region ($\alpha$$>$0.9), while the
intermediate energy DAR-$\pi$ $\nu$'s allow
measurements in the transition regions (0.9$>$$\alpha$$>$0.1).


\begin{table}
\caption{
Maximum neutrino energy ($\Enu$)
with which coherency is maintained
among the constituents,
as characterized by the
parameters $F ( \q2 _{max} )$, $\alpha$ and $\xi$
being $>$0.95.
}
\begin{ruledtabular}
\begin{tabular}{cccc}
Parameter & \multicolumn{3}{c}{Maximum $\Enu$(MeV) for}\\
$>$ 0.95  & Ar & Ge & Xe \\ \hline
$F ( \q2 _{max} )$ & 17.2 & 14.1  & 11.6 \\
$\alpha$ at $\T0$=0 & 21.1 & 17.4 & 14.3 \\
$\xi$ at $\T0$=0 & 21.6 & 17.6 & 14.4
\end{tabular}
\end{ruledtabular}
\label{tab::Enu95}
\end{table}


An objective with the studies
of $\nuA$ scattering is 
to probe physics beyond SM~\cite{nuA-BSM,scholberg}. 
A direct approach would be to compare the 
measured cross-sections with the
SM predictions given in 
Eqs.~\ref{eq::diffsigmanuA}\&\ref{eq::totalsigmanuA}.
The sensitivities would be limited 
by the uncertainties of the 
form factors in describing the nuclear effects.
This favors measurements to be performed
at regimes of $F ( \q2 )$$\simeq$1.
The maximum $\Enu$'s which retain 
$F ( \q2 _{max} )$$>$0.95   
are listed in Table~\ref{tab::Enu95}.

Another distinctive BSM signature 
is that $\sigmanuA$ would no longer
vary as $[ \varepsilon  Z - N ] ^2$
even in the coherency regime of $\alpha$$\simeq$1.
Dependence characteristics 
of the deviations can reveal the nature 
of the new physics couplings.
For instance, anomalous neutrino
magnetic moments would
give rise to an additional contribution 
which scales as $Z^2$
(that is,  
$( \mathcal{X}_j , \mathcal{Y}_k )$=(1,0)
in Eq.~\ref{eq::amplitude})~\cite{NMMVE89},
while BSM physics
giving rise to the dark matter spin-independent 
couplings are usually taken as
varying with $A^2$ or
$( \mathcal{X}_j , \mathcal{Y}_k )$=(1,1)~\cite{rppdarkmatter}.
Sensitivities would be constrained
for measurements in kinematics space 
where coherency is partial.
The maximum $\Enu$
at $\T0$=0 which maintain coherency
with $\alpha$ and $\xi$ at  $>$0.95 
are shown in Table~\ref{tab::Enu95}.
Low energy neutrino sources like reactor 
and solar neutrinos are better suited 
to probe BSM effects in $\nuA$, 
although more sensitive experiments 
are necessary since the measurable recoil energy is lower. 
These experiments do not have beam-structures 
for background subtraction but are not vulnerable to 
neutrino-induced neutron background like 
those with DAR-$\pi$ sources.
 
Detailed quantitative studies on
the search strategies and
potential reaches of different BSM models with 
$\nuA$ interactions, as well as the 
sensitivity constraints
due to decoherency effects
are subjects of future research.

This work is supported by contracts 
102-2112-M-002-013-MY3,
104-2112-M-001-038-MY3 and
104-2112-M-259-004-MY3
from the Ministry of Science and Technology, Taiwan,
2015-ECP4 from 
the National Center of Theoretical Sciences, Taiwan,
114F374 from TUBITAK, Turkey, and
11475117 from the National Natural Science Foundation, China.

\end{document}